\documentclass[twocolumn]{aastex631}

\newcommand{\pcm}{\,cm$^{-2}$}	
\newcommand{\psec}{\,s$^{-1}$}	
\newcommand{\src}{J204734}	
\newcommand{\erg}{\,erg cm$^{-2}$ s$^{-1}$}	
\newcommand{\lum}{\,erg s$^{-1}$}	

\newcommand{\xmm}{\emph{XMM-Newton}}	
\newcommand{\chandra}{\emph{Chandra}}	

\begin{document}

\title{Intriguing nature of AM Her type candidate CXOU J204734.8+300105}

\correspondingauthor{Rahul Sharma}
\email{rahul1607kumar@gmail.com}

\author[0000-0003-0366-047X]{Rahul Sharma}
\affiliation{Raman Research Institute, C.V. Raman Avenue, Sadashivanagar, Bengaluru 560080, INDIA}
\affiliation{Inter-University Centre for Astronomy and Astrophysics (IUCAA), Ganeshkhind, Pune 411007, INDIA}

\author[0000-0002-5562-7965]{Chetana Jain}
\affiliation{Hansraj College, University of Delhi, Delhi 110007, INDIA}


\author{Biswajit Paul}
\affiliation{Raman Research Institute, C.V. Raman Avenue, Sadashivanagar, Bengaluru 560080, INDIA}

\author{Anjan Dutta}
\affiliation{Department of Physics \& Astrophysics, University of Delhi, Delhi 110007, INDIA}

\author[0000-0003-1703-8796]{Vikram Rana}
\affiliation{Raman Research Institute, C.V. Raman Avenue, Sadashivanagar, Bengaluru 560080, INDIA}



\begin{abstract}

The detection and characterization of periodic X-ray signals are crucial for identifying new compact objects and studying the mechanisms powering their emission. We report on the timing and spectral variability of CXOU J204734.8+300105, a proposed eclipsing polar-type cataclysmic variable (CV) candidate. This source has been observed once with \chandra\ and twice with \xmm, revealing several intriguing and conflicting features in its X-ray emission. The \chandra\ observation showed a periodicity of $\sim$6000 s with an eclipse-like feature. The X-ray light curve from 2017 \xmm\ observation showed a period of $\sim$2000 seconds without any apparent eclipse, while the simultaneous optical light curve from OM showed a period of $\sim$6000 seconds. This variability raises questions about the true nature of the source. Spectral analysis indicates a multi-component emission and emission lines due to Fe. The spectral characteristics are consistent with those observed in other CV systems. Additionally, we identified optical and near-infrared counterparts from various catalogues. Our findings suggest a dynamic and evolving accretion environment of CXOU J204734.8+300105.

\end{abstract}



\section{Introduction} \label{sec:intro}

Cataclysmic variables (CVs) are compact binary star systems consisting of a white dwarf accreting material from a late-type companion star \citep{Hellier01, Warner}. The orbital periods of these close binaries range from about 80 mins to several hours \citep{Knigge11}. These variable stars are categorized into various sub-types based on photometric variability, spectroscopic and polarimetry characteristics in the optical and infrared, and timing and spectroscopic properties of their X-ray emission.

The accretion onto the white dwarfs is strongly affected by the strength of their magnetic field. CVs can be classified into two major classes based on the strength of the magnetic field of the white dwarf primary: non-magnetic ($\lesssim$$10^5$ G) and magnetic ($\gtrsim$$10^6$ G) CVs. In non-magnetic CVs, the magnetic field is not strong enough to affect the accretion flow. Accretion occurs from a surrounding accretion disk that extends to the surface of the white dwarf. In magnetic CVs, the accretion flow is governed by the magnetic field, which channels matter along the magnetic field lines onto the magnetic poles of the white dwarf, releasing gravitational energy as X-ray emission. 
The latter is further divided into two major subclasses: Polars and Intermediate Polars. In intermediate polars \citep[or DQ Her type systems;][]{Patterson94}, a partial accretion disk is usually present, truncated by the magnetosphere of the white dwarf, and the spin period ($P_{\rm spin}$) of the white dwarf is significantly shorter than the binary orbital period ($P_{\rm orbital}$). In polars \citep[or AM Her type systems;][]{Cropper90}, the magnetic field is strong enough to prevent the formation of an accretion disk. In these systems, the accretion region emits cyclotron radiation, contributing significantly to the infrared and optical energy bands, which are strongly polarized. The magnetic torque in these systems results in the white dwarf rotating synchronously with the orbital period ($P_{\rm spin}$=$P_{\rm orbital}$). Some polars, known as asynchronous polars, exhibit a slight mismatch between $P_{\rm spin}$ and $P_{\rm orbital}$ by $<1$\%.


The present study investigates the X-ray variability exhibited by the polar type candidate, CXOU J204734.8+300105 (hereafter \src), located in the Cygnus loop, which was discovered during the \chandra\ ACIS Timing Survey Project \citep{Garmire03, Israel16}. Based on a periodicity of $\sim 6097$ s from \chandra\ data and the presence of an eclipse-like feature, \citet{Israel16} categorized this source as an eclipsing polar-type CV. The optical counterpart of this source shows the presence of H and He emission lines. Additionally, \citet{Farrell2015} and \citet{Yang2022} classified this source as a CV candidate using a machine learning algorithm based on the \xmm\ and \chandra\ data, respectively. 
In this work, we have explored the timing and spectral variability in \src\ using \xmm\ and \chandra\ observations in order to understand the emission mechanism from this interesting class of galactic X-ray sources.

\section{Observations \& Data Analysis}

In this work, we have analyzed the pointed X-ray observations of \src\ made with the instruments onboard the \chandra\ and \xmm\ missions. The observation details are given in Table~\ref{table:log}. In addition, we have also incorporated optical data from the Optical Monitor (OM) onboard \xmm\ and the Zwicky Transient Facility (\emph{ZTF}) to investigate the multi-wavelength variability of the source. 

\begin{figure*}
	\includegraphics[width=0.459\linewidth]{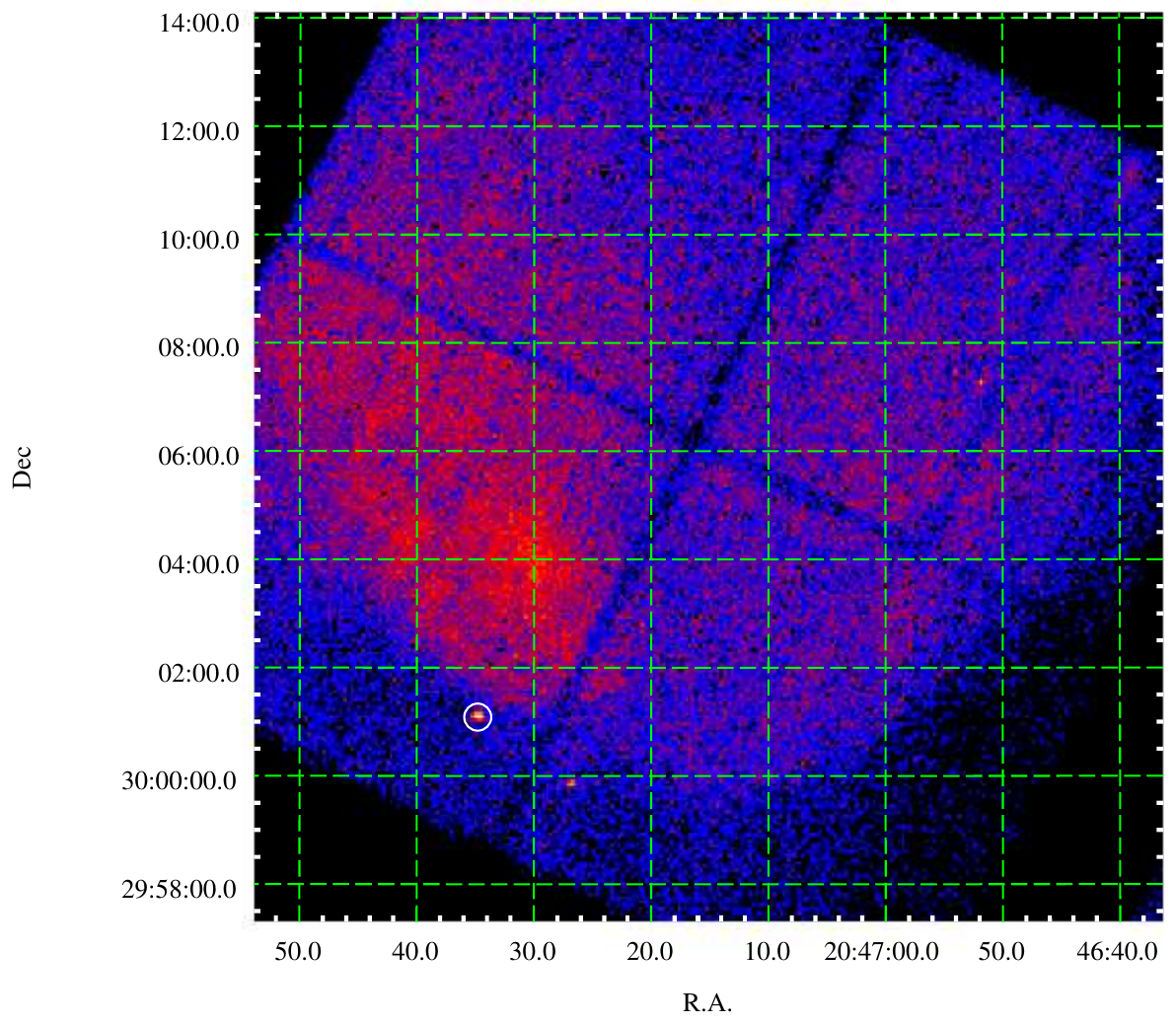}
    \includegraphics[width=0.459\linewidth]{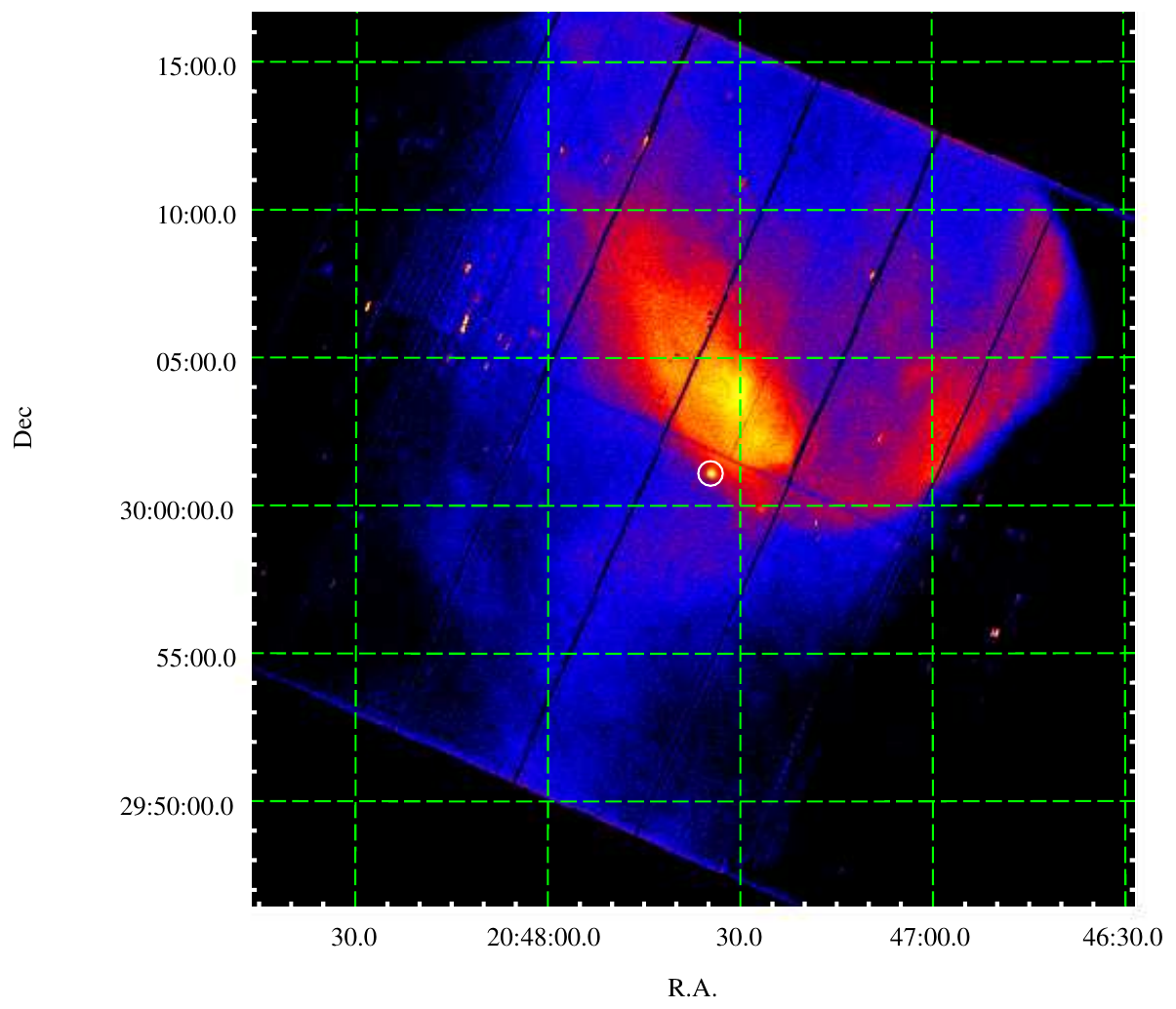}
    \caption{The DS9 image of \src\ from \chandra\ (\textit{Left Panel}) and \xmm-EPIC-PN observation of 2017 (\textit{Right Panel}). The source region of 15\arcsec is marked in a white circle. }
    \label{fig:image}
\end{figure*}

\begin{figure*}
\centering
    \includegraphics[width=0.53\linewidth]{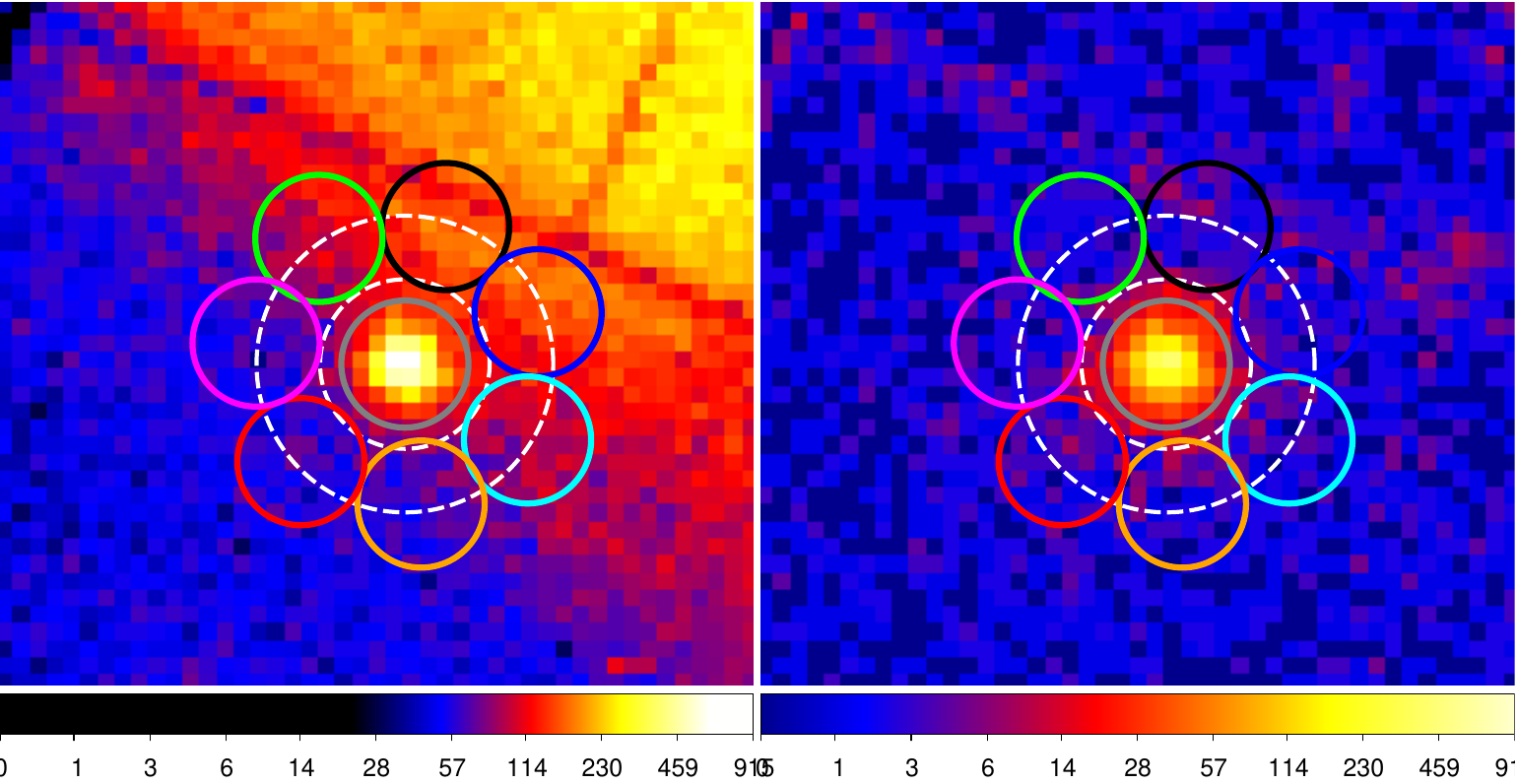}
    \includegraphics[width=0.459\linewidth, height=0.22\textheight]{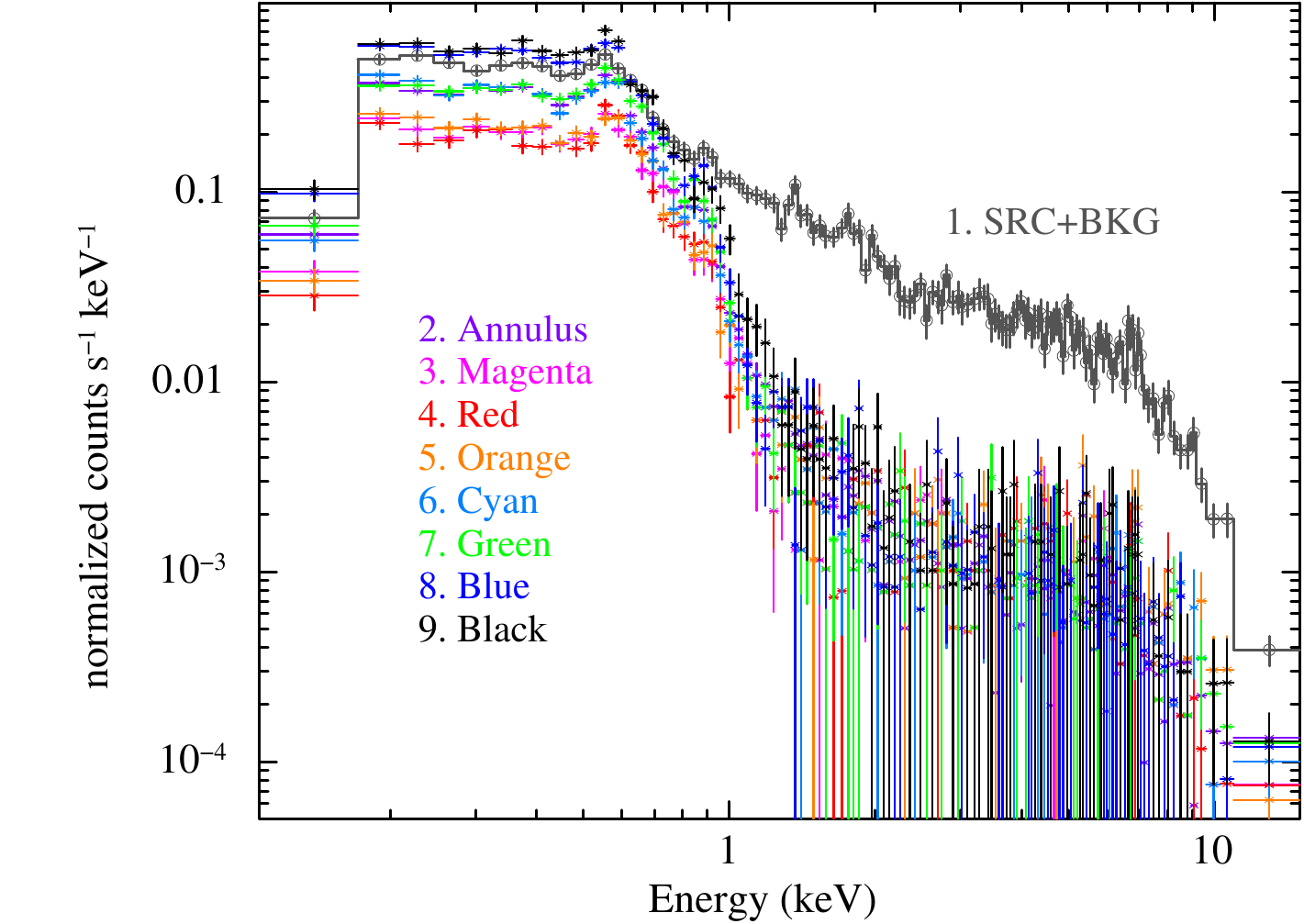}
    \caption{ \textit{Left}: Image of the source from the 2017 \xmm\ observation. The left panel shows the image without any energy selection, while the right panel is energy-selected within the 1.5--10 keV range.
    The source region, marked with a grey circle, has a radius of 15\arcsec. Several similar-sized circular regions around the source represent different background regions. An annular region is indicated with two dashed white circles, having an inner radius of 20\arcsec and an outer radius of 35\arcsec. 
    \textit{Right}: Comparison of the source and background spectra for the 2017 \xmm\ observation. Different colors correspond to the various background regions shown in the left image. }
    \label{fig:srcbkg}
\end{figure*}

\subsection{\chandra}

\src\ was observed with Chandra X-ray Observatory \citep{Weisskopf00} in 2000 May for an exposure time of about 10 ks with Advanced CCD Imaging Spectrometer \citep[ACIS;][]{Garmire03} detector. We used the Chandra Interactive Analysis of Observations \citep[CIAO;][]{Fruscione2006} software (v. 4.14; Caldb v. 4.9.8) for data reduction. A level 2 data set was produced using the \textit{chandra\_repro }script. The left panel of Figure ~\ref{fig:image} shows an image extracted from \chandra\ ACIS detector. The source is marked with a 15\arcsec\ circular region at the source location RA= 20$^h$ 47$^m$ 34.8$^s$ and DEC=+30$^\circ$ 01$'$ 05.2\arcsec\ \citep{Israel16}. An annular region of an inner (outer) radius of 20\arcsec\ (35\arcsec) was chosen to make a background region around the source (see sec. \ref{sec:xmm}). The background subtracted and barycenter corrected light curves were obtained using the CIAO-tool \texttt{dmextract} and \texttt{axbary}, respectively. The source and background spectra were created with the \texttt{specextract} task. The estimated pile-up fraction during this observation was about 8\% and can be neglected for the scientific objectives of this work. 

\begin{table}
	\centering
	\caption{The X-ray observation log of \src.}
    \label{table:log}
	\begin{tabular}{lccc} 
		\hline
		Instrument & Obs-ID & Obs. Date & Exposure (ks)\\
		\hline
		\chandra & 740 & 2000-05-21 & 9.8 \\
		\xmm & 0082540701 & 2002-12-13 & 13.9 \\
		\xmm & 0803660101 & 2017-11-20 & 26.7 \\
		\hline
	\end{tabular}
\end{table}

\subsection{\xmm}
\label{sec:xmm}
\src\ was observed twice with \xmm\ \citep{Jansen2001} in full frame mode, once in 2002 and then in 2017 (see Table \ref{table:log}). In the 2002 dataset, the source was detected near the corner of EPIC CCDs (off-axis)
, while in the 2017 dataset, the source was observed at the centre of the CCDs (on-axis). 
The \xmm\ Science Analysis Software (SAS, v.20) was used to process the data from the European Photon Imaging Camera (EPIC), which is equipped with PN \citep{Struder01} and MOS-type \citep{Turner01} charge-coupled device (CCD) detectors. The images, spectra and light curves were processed from EPIC event files by using the SAS task \texttt{evselect}. Similar to the \chandra\ data, the right panel of Fig.~\ref{fig:image} shows the selection of the source \src\ for \xmm\ data of 2017 with EPIC-PN. 

The source is situated within the Cygnus Loop remnant, which is a strong source of soft X-rays \citep{Fesen1982}. This poses challenges in selecting an appropriate background region. To address this issue, we adopted different circular regions with a radius of 15\arcsec\ around the source as background, along with an annulus region of the inner and outer radius of 20\arcsec\ and 35\arcsec, respectively (as illustrated in the left panel of Fig. \ref{fig:srcbkg} with different colors). The right panel of Fig. \ref{fig:srcbkg} presents a comparison of spectra obtained from the source and background regions, with different colors indicating distinct background selections. Below 0.9 keV, the background spectrum displayed considerable variation depending on the chosen region. Above 0.9 keV, the background spectra exhibited consistency across different selections; the same can be seen with the \xmm\ image in 1.5--10 keV (Fig. \ref{fig:srcbkg}). Consequently, we opted for the annulus region for background selection in subsequent analysis. We also adopt the same annular background region for \chandra\ data analysis for the same reason. Notably, the source counts substantially exceeded the background within the 0.9--10 keV energy range, therefore, this energy range was chosen for further spectral analysis.

The \texttt{barycen} and \texttt{epiclccorr} tasks were applied to extract the light curves. The former converts the photon arrival time into the solar system barycenter time, and the latter performs a series of corrections to minimize effects that may impact the detection efficiency (such as dead time, chip gaps, and point-spread-function variation) before producing a background-subtracted light curve. We summed the light curves of the individual MOS detector cameras to obtain a cumulative light curve using \texttt{lcmath}. The ancillary response files and the spectral redistribution matrices for the spectral analysis were generated with \texttt{arfgen} and \texttt{rmfgen}, respectively. Each of the three spectra from the EPIC detectors was rebinned with a significance of at least 3$\sigma$ for each energy bin, using the SAS tool \texttt{specgroup}. No pile-up was observed in either of the observations.

For the 2017 observation, the OM \citep{XMM-OM} was simultaneously operated in imaging fast mode with the U filter (300-390 nm) for 5 exposures of each 4380 s. The optical light curves were extracted from the pipeline-processed products (PPS) and were merged for all exposures after barycentric correction. The average net count rate for OM data is $\sim 2.4$ counts \psec, equivalent to an instrumental magnitude of $\sim 18.2$ mag\footnote{The zero-point value of 19.189 for the U filter was used following the AB-magnitude system \url{https://www.cosmos.esa.int/web/xmm-newton/sas-watchout-uvflux}.}.

\subsection{Zwicky Transient Facility}
\emph{ZTF} \citep{Bellm2019} is an optical wide-field survey project using a CCD camera array attached to the Samuel Oschin telescope at Palomar Observatory in California, operated by the California Institute of Technology (Caltech). We utilized $g$ and $r$ band light curves of the targets acquired from \emph{ZTF} Public Data Release 21 \citep[DR21;][]{Masci2019}, which covers the epoch of $\sim$58206--60368 MJD. The $g$ and $r$ bands have 550 and 1133 measurements, respectively. 

\section{Results}

\begin{figure}
	\includegraphics[width=0.9\columnwidth]{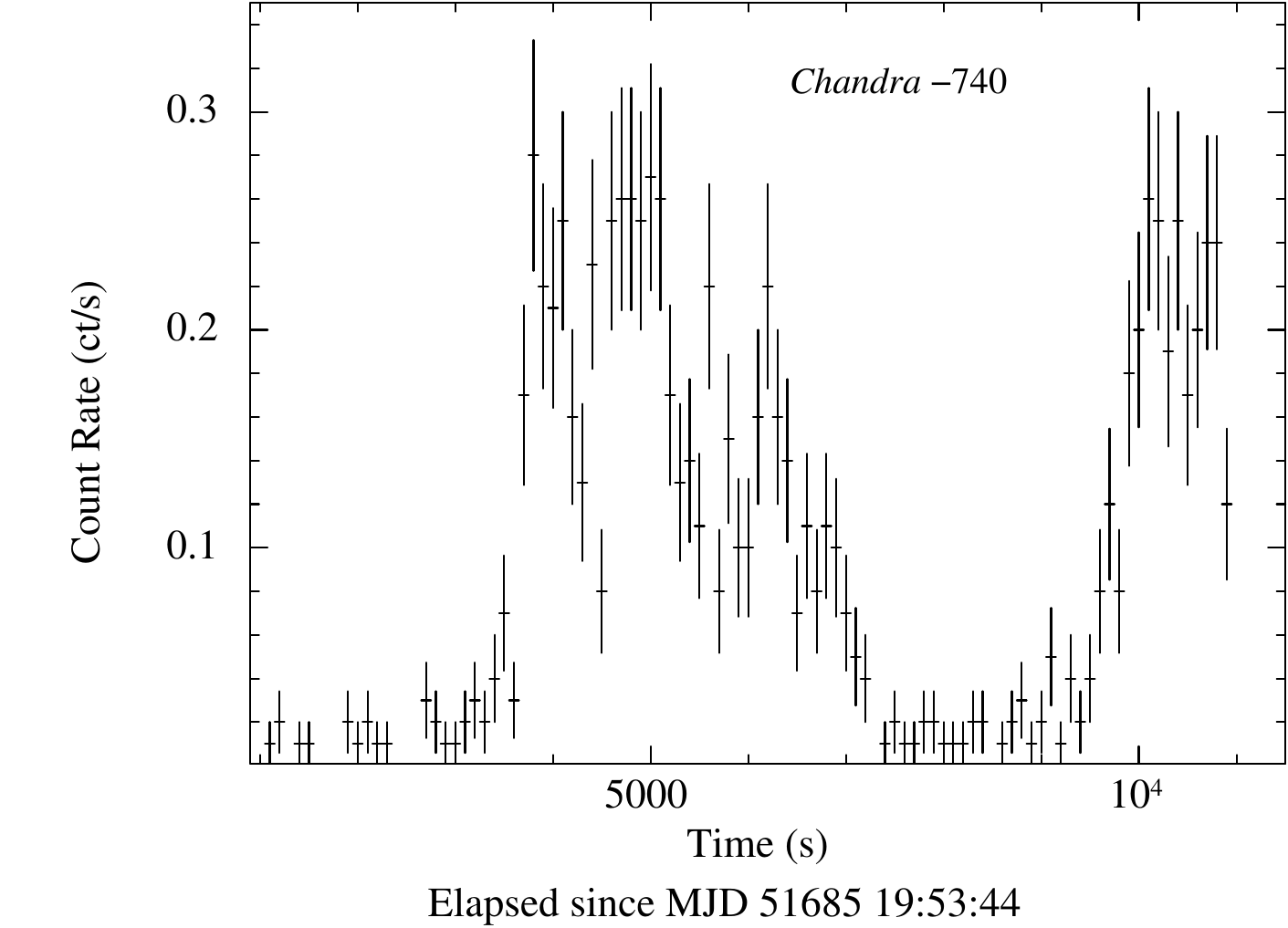}
    \caption{The 0.5--10 keV \chandra\ light curve of \src\ binned at 100 s, clearly showing an orbital period of about 6000 s, including an X-ray eclipse-like feature lasting for $\sim$2000 s.}
    \label{fig:chandralc}
\end{figure}

\begin{figure*}
\centering
    \includegraphics[width=0.9\linewidth]{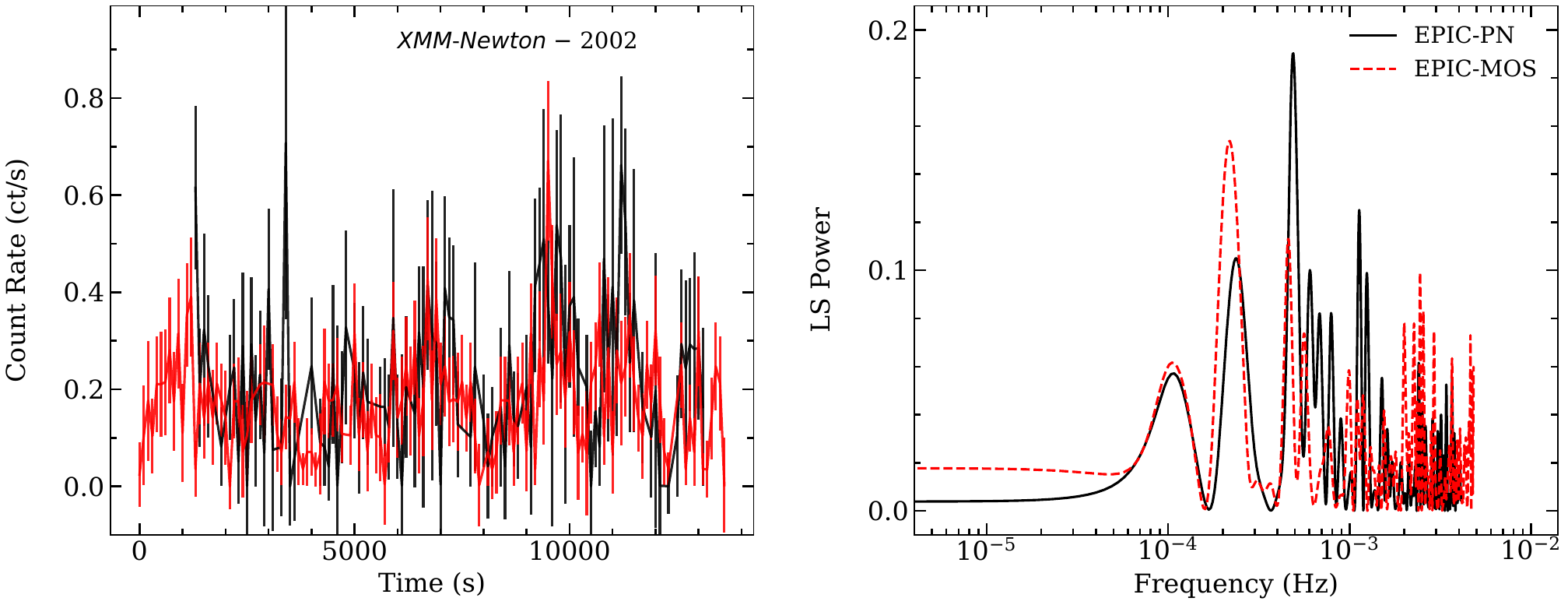}
    \includegraphics[width=0.9\linewidth]{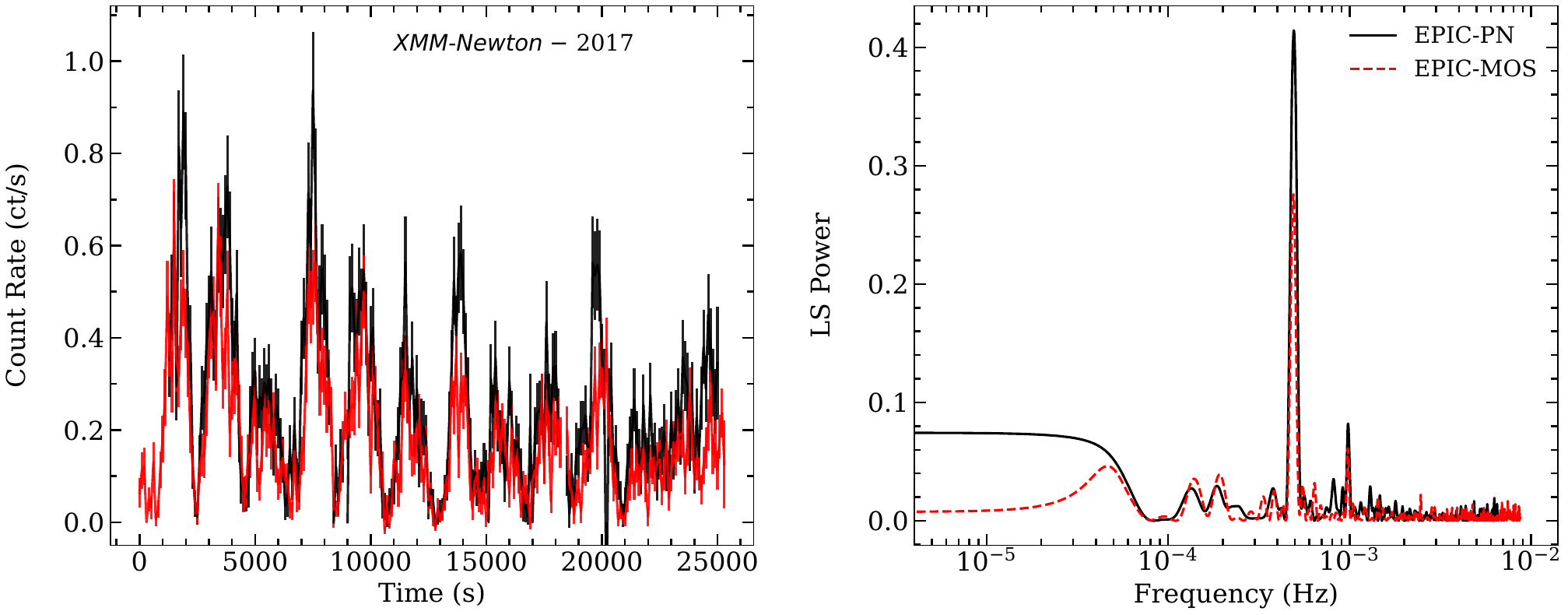} 
        \caption{\textit{Left panel}: The EPIC-PN and MOS light curve of \src\ from 2002 (top) and 2017 (bottom) \xmm\ observations binned at 100 s in the energy range of 1--10 keV. \textit{Right panel}: The LS periodogram of \src\ generated from \xmm-EPIC-PN and MOS light curves. The periodogram power has been normalized by the residuals of the data around the constant reference. A marginally significant peak is observed in the periodogram at $\sim$2043 s in PN data only of the 2002 observation, while strong coherent pulsations at a similar period were observed in PN and MOS data of the 2017 observation.}
    \label{fig:xmmlc}
\end{figure*}

\subsection{Timing Analysis}

Fig. ~\ref{fig:chandralc} shows the 0.5--10 keV \chandra\ light curve of \src\ binned at 100 s. It reveals an eclipse-like feature lasting approximately 2000 s and an orbital period of about 6000 s \citep{Israel16}. However, subsequent \xmm\ observations do not exhibit this behaviour. Fig. ~\ref{fig:xmmlc} presents the 1--10 keV \xmm\ light curves for 2002 (top left) and 2017 (bottom left) observations. The 2002 observation does not show any clear modulations, whereas the 2017 observation reveals modulations at $\sim$2000 s,  evident in both the PN and MOS light curves. 

To investigate these modulations, we used a Lomb-Scargle (LS) periodogram \citep{Lomb1976, Scargle1982}. We used the Lomb-Scargle periodogram package \citep{VanderPlas18} from \textsc{astropy} \citep{astropy}, and 20 ‘samples per peak’ as the oversampling. The right panels of Fig.~\ref{fig:xmmlc} present the LS periodogram for 2002 (top) and 2017 (bottom) \xmm\ observations, with power normalized by the residuals of the data around the constant. 
The periodogram of 2017 observations shows a highly significant peak corresponding to a periodicity of 0.5 mHz in both the PN and MOS datasets. The peak observed with the PN data was highly significant, with a false alarm probability of $<10^{-10}$ as estimated using the bootstrap method. A harmonically related peak at $\sim$1 mHz is also evident. In contrast, the 2002 periodogram shows a weaker peak at 0.5 mHz, detected only in the PN data, and with much lower significance (false alarm probability = 0.3). No corresponding peak is detected in the MOS data. Furthermore, no significant power is observed around a period of $\sim$6000 s in either dataset.


To find the uncertainties of the peak of the LS periodogram, we employed bootstrapping, simulating light curves based on the original data following \citet{Boldin13} and \citet{Sharma23}. We performed 10,000 simulations and identified the peak of each LS periodogram. The standard deviation on the observed peak period from the simulated light curve was taken as the 1$\sigma$ uncertainty. To refine the period of modulation for the 2017 observations, we also used the epoch-folding method \citep{Leahy1983}. The uncertainty on period was determined using the bootstrap method. Table~\ref{tab:period} lists the periodicity with errors obtained from both techniques. Although the significance of peak for the 2002 observation is low, we found the peak around a similar period of 2043 s with 1$\sigma$ uncertainty of 43 s. 

\begin{figure}
	\includegraphics[width=0.9\columnwidth]{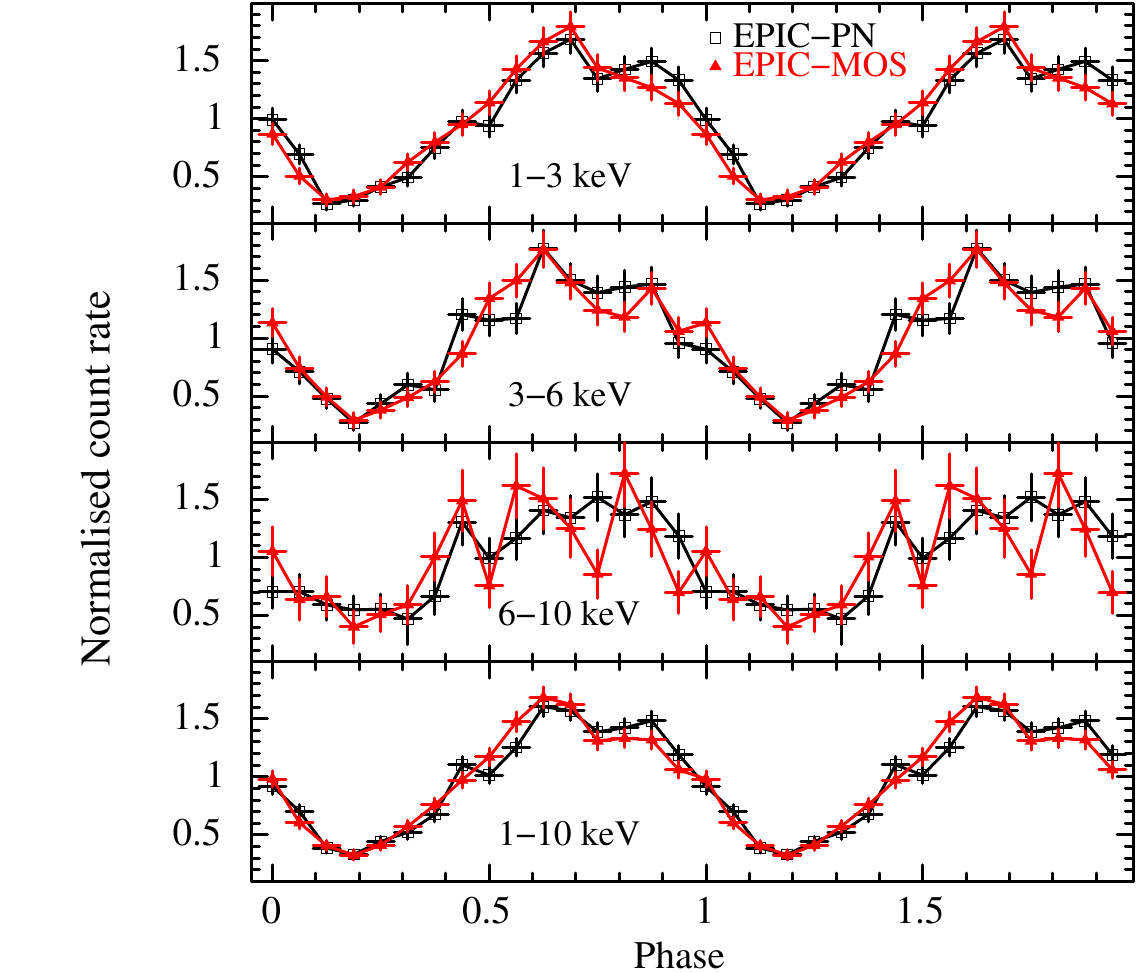}
    \caption{The energy-resolved pulse profiles of \src\ from the 2017 \xmm\ EPIC-PN and MOS data in the energy ranges of 1--3, 3--6, 6--10 and 1--10 keV. The light curves were folded using the detected periodicity from the epoch-folding method, as listed in Table~\ref{tab:period}.}
    \label{fig:pulseprofile}
\end{figure}

Fig. ~\ref{fig:pulseprofile} shows the energy-resolved EPIC-PN and MOS pulse profiles of \src\ from the 2017 observation. The light curves were folded using the period obtained with the epoch-folding method as listed in Table~\ref{tab:period}. From top to bottom, the pulse profiles are shown in the energy range of 1--3, 3--6, 6--10, and 1--10 keV. The pulse profiles exhibit an asymmetric sinusoidal shape with a pulse fraction of up to $\sim$70\% in the 1--10 keV range.  The pulsed fraction was observed to decrease with increasing energy.

\begin{table}
	\centering
	\caption{The periodicity (in seconds) obtained using the Lomb-Scargle periodogram and epoch-folding method applied on the 2017 \xmm\ observation of \src. The numbers in the brackets indicate a 1$\sigma$ error in period determination.}
	\label{tab:period}   
	\begin{tabular}{lcc} 
		\hline
		Instrument & LS & Epoch folding\\
		\hline
		EPIC-PN & 2030$\pm$7 & 2027$\pm$3  \\
		EPIC-MOS & 2042$\pm$7 & 2045$\pm$3  \\
		\hline
	\end{tabular}
\end{table}

\subsection{Spectral Analysis}



For spectral analysis, we used \textsc{xspec} version 12.9.1 \citep{Arnaud}. For spectral fitting, \texttt{tbabs} was used to model interstellar absorption with elemental abundances set to \texttt{wilm} \citep{Wilms} and photoelectric absorption cross-sections of \citet{Verner}. For \xmm\ observations, the spectrum from EPIC-PN, MOS1 and MOS2 detectors were fitted simultaneously, and a multiplicative constant was used to account for the cross-calibration uncertainty of the instruments. The constant was fixed to 1 for the PN spectrum and allowed to vary for MOS spectra. All the spectral uncertainties and the upper limits reported in this paper are at a 90\% confidence level unless specified.

The spectra obtained from \chandra\ (the whole observation) and the first \xmm\ were of poor statistics and were well described with an absorbed powerlaw model. The resulting fits yielded $\chi^2$/dof values of 26.3/34 and 21/25, respectively. The photon indices were $0.38 \pm 0.11$ for \chandra\ and $0.67^{+0.23}_{-0.13}$ for the \xmm\ observation, indicating a relatively harder spectrum in the \chandra\ data. Additionally, the source exhibited a higher flux during the \chandra\ observation, measured at $2.6 \times 10^{-12}$ \erg, compared to $1.1 \times 10^{-12}$ \erg\ during the first \xmm\ observation, both in the 0.9--10 keV energy range. 


The 2017 spectrum was distinctly different from previous observations of \src\ and had enough statistics to perform a detailed spectral analysis. We used the absorbed powerlaw model to describe the 0.9--10 keV source continuum spectrum, which resulted in an unsatisfactory fit ($\chi^2$/dof $\sim$260.7/221), leaving residuals at soft X-ray energies and 6--7 keV. The residuals at 6-7 keV hint at the iron emission line and were modelled using a \texttt{Gaussian} emission component. The residuals at soft X-ray can be addressed by the addition of either a soft thermal blackbody component (\texttt{bbodyrad}) or a partial covering absorber (\texttt{tbpcf} in \textsc{xspec}). 
The added blackbody or the partial covering absorption led to a significantly improved description of the spectra below 1.5 keV, improving the reduced chi-squared to $\chi^2_{\nu} \sim 1$. The additional components (blackbody or partial covering absorber and a Gaussian emission line) were significant at more than 3$\sigma$ with the F-test probability of $\sim$10$^{-7}$. We also assessed the likelihood of these features and performed $10^4$ Monte Carlo simulations using the \textsc{xspec} routine \texttt{simftest}. The simulation of the \xmm\ spectra revealed a high significance of blackbody/partial absorber and Gaussian emission line with a probability of $\sim5\times10^{-4}$ and $<10^{-4}$, respectively. 

\begin{figure}
	\includegraphics[width=\columnwidth]{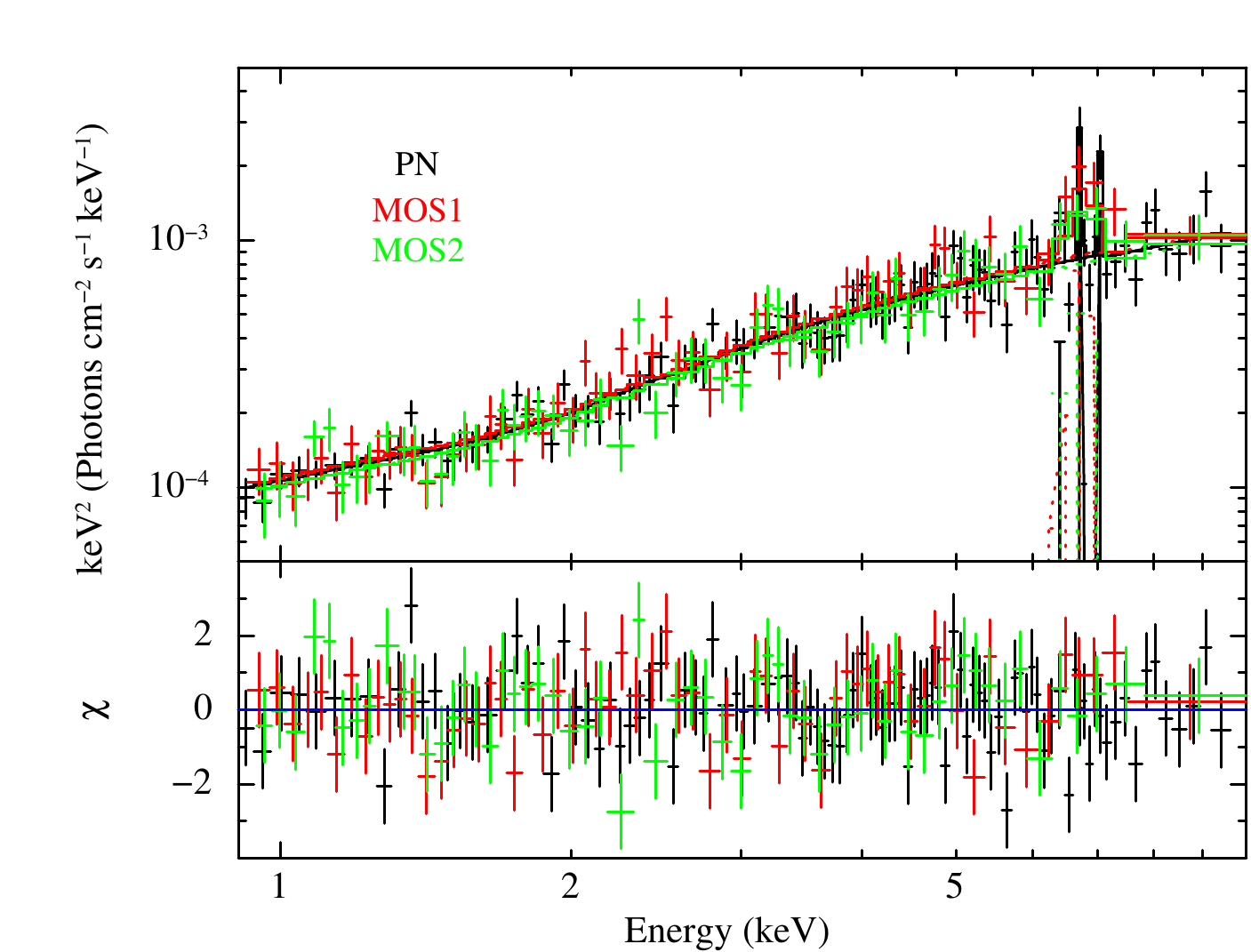}
    \caption{Phase-averaged spectra of \src\ modelled with partially absorbed powerlaw and three Gaussian emission components. The lower panel represents the residual from the best-fit spectral model.}
    \label{fig:spec}
\end{figure}

The blackbody component provided a temperature of $\sim$0.13 keV, while the partial absorber showed a column density of $\sim$$7-8 \times10^{22}$ \pcm\ with a covering fraction of 54\%. The power-law component associated with these models had a photon index of $\sim$1.3. For a model with a soft blackbody component, normalization of the blackbody component implies a compact emission region with a radius of $\sim$9 km, likely originating from the surface of the white dwarf.
The Gaussian emission line was observed to be around 6.8 keV with a width of 0.2 keV, suggesting ionized Fe line emission possibly blends from multiple ionizations. Therefore, we replaced the Gaussian emission line with three Gaussian lines at 6.4, 6.7 and 6.97 keV for neutral, He-like and H-like Fe, respectively.  
The width of these Gaussian emission lines was fixed at 10 eV. All three Gaussian line energies were consistent with the assumption and were detected at 6.39 (10), 6.73 (6) and 7.02 (8) keV with equivalent widths of $72^{+71}_{+68}$, $144^{+107}_{-64}$ and $122^{+106}_{-76}$ eV, respectively. 
\src\ spectra with a best-fit spectral model comprising partially absorbed powerlaw components along with three Gaussian emission line components corresponding to the neutral, H-like and He-like Fe K$_\alpha$ lines are shown in Fig. ~\ref{fig:spec}. 

We also checked other continuum models, such as thermal bremsstrahlung (\texttt{bremss}), collisionally ionized gas (\texttt{APEC}) and multi-temperature cooling-flow model \citep[\texttt{mkcflow};][]{Mushotzky88} models instead of powerlaw. The partially covered bremsstrahlung model with a Gaussian emission line could well fit the spectra ($\chi^2$/dof=213/216), but the temperature could not be constrained and gives a lower limit on temperature of 25 keV. Similarly, \texttt{mkcflow} with Solar abundance was able to describe the continuum along with accounting for the Fe emission line but was not able to constrain the lower and upper temperature and give lower and upper limits of 10 keV and 22 keV, respectively. 

\begin{figure*}
\centering
    \includegraphics[width=0.69\columnwidth]{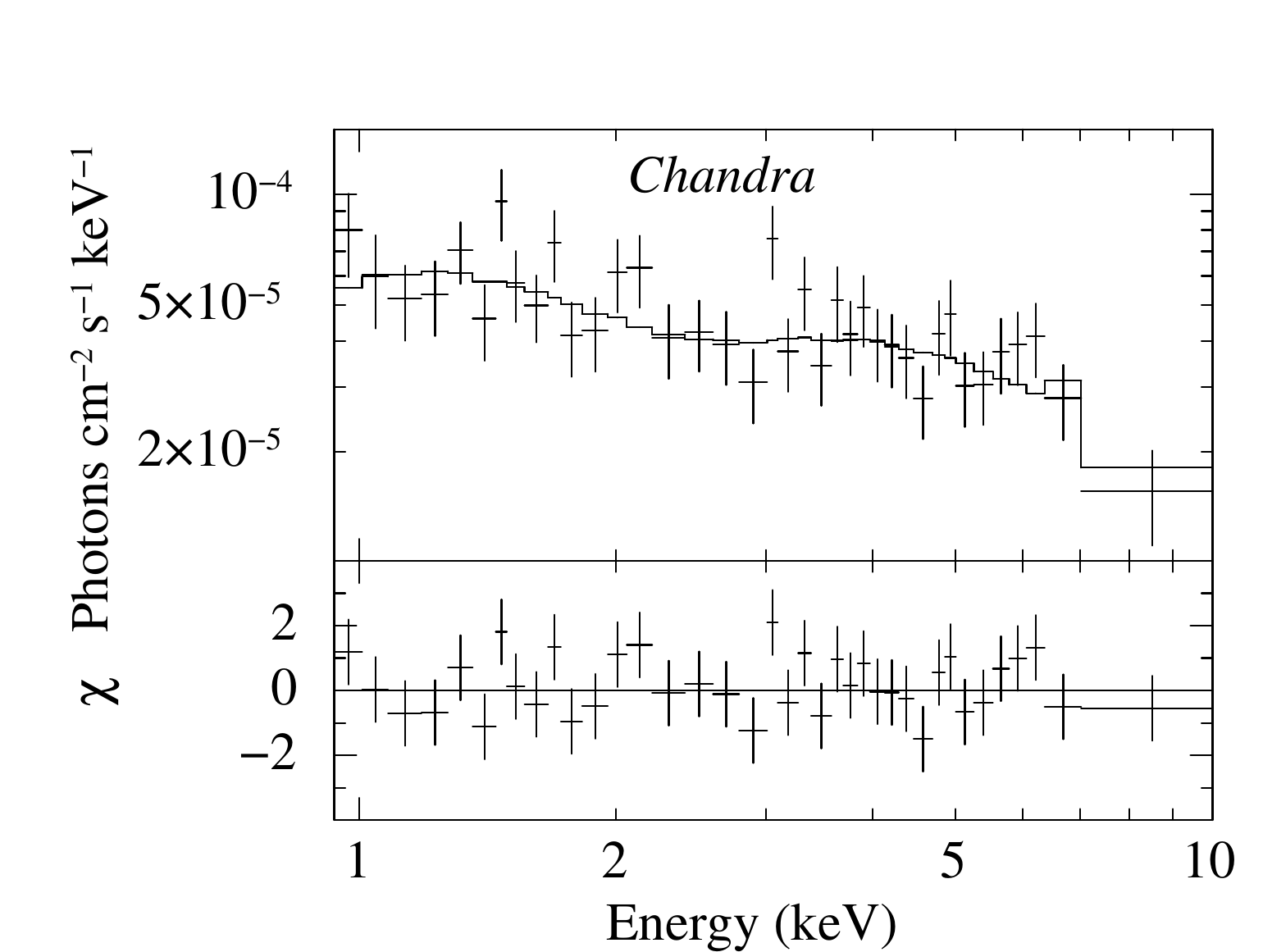}
    \includegraphics[width=0.687\columnwidth]{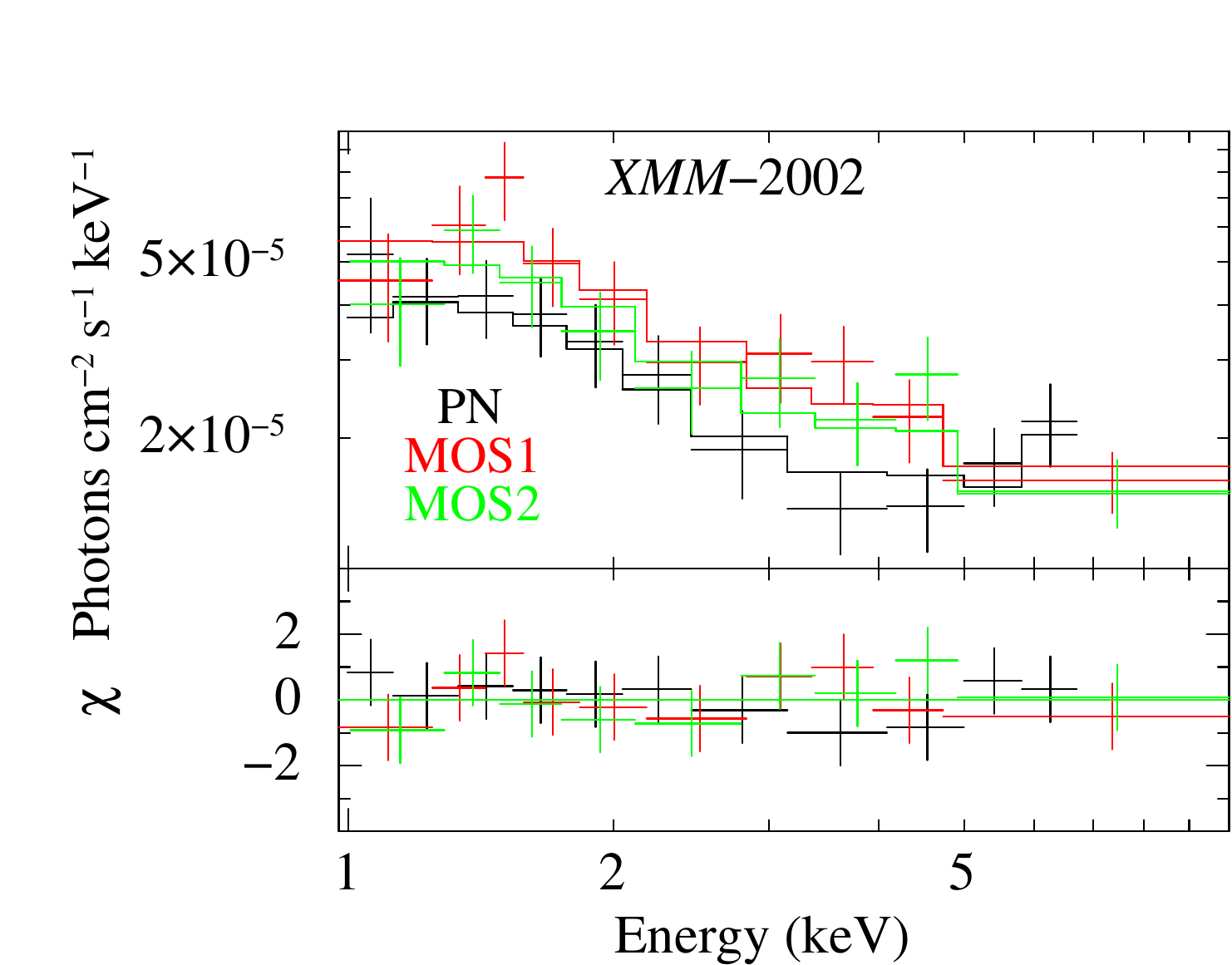}
    \includegraphics[width=0.687\columnwidth]{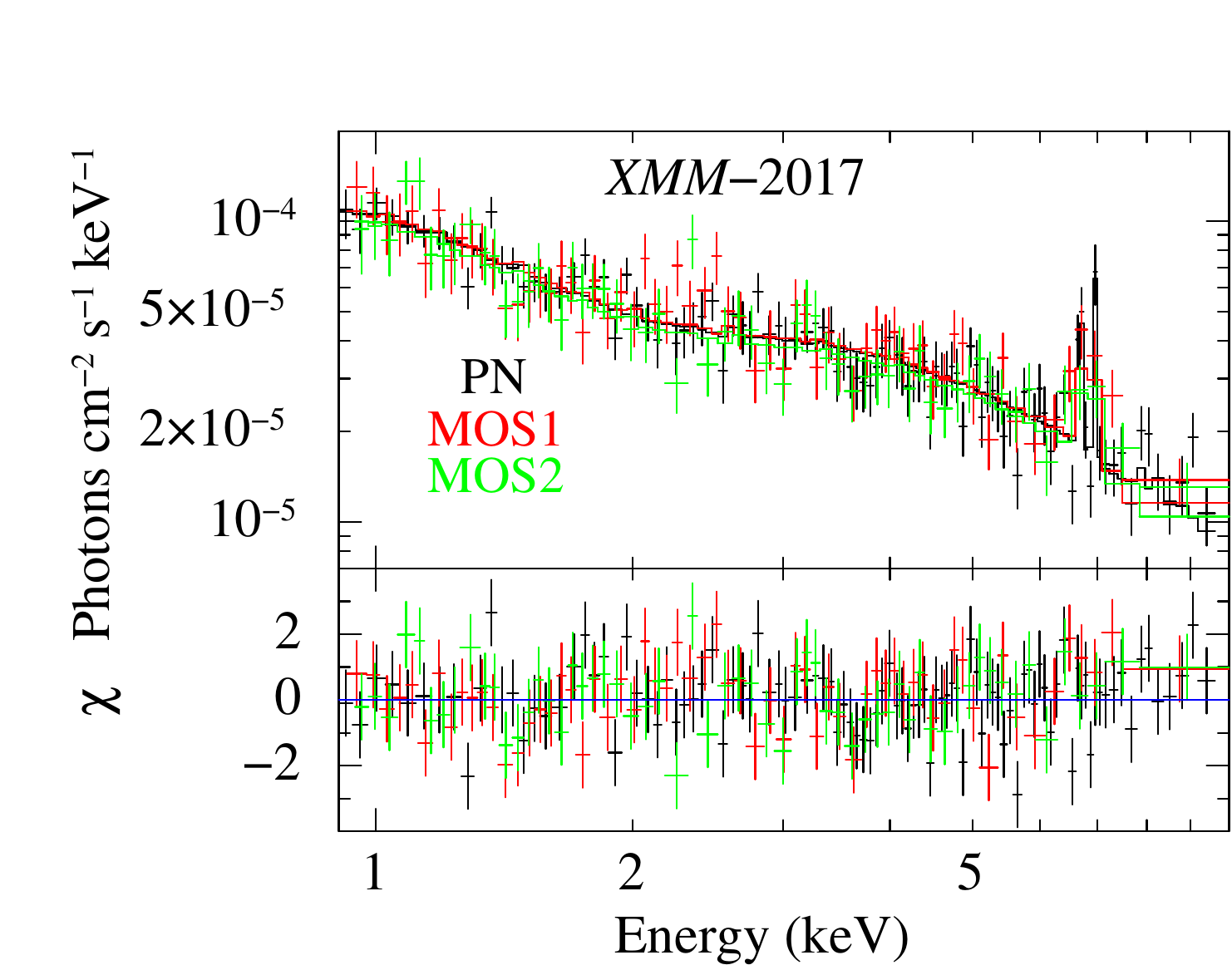}
    \caption{The \chandra\ (left), 2002 \xmm\ (middle) and 2017 \xmm\ (right) spectrum modelled with partially absorbed \texttt{APEC} model. The lower panels represent the residual from the best-fit spectral model.}
    \label{fig:allspec}
\end{figure*}

The partially covered \texttt{APEC} model could also fully describe the continuum and account for the ionized Fe emission line feature. The \texttt{APEC} plasma has a temperature of $\sim 11.7$ keV, with abundance assumed to be Solar and fixed to 1. The partial absorber suggested a larger column density compared to that required by the powerlaw continuum. Although this model gave slightly poorer fit statistics $\chi^2$/dof$\sim$223.8/219 relative to the partially absorbed power-law model ($\chi^2$/dof$\sim$213/219), the \texttt{APEC} model is more physically motivated given the thermal nature of the emission. We also applied the partially absorbed \texttt{APEC} model to the \chandra\ and 2002 \xmm\ observations. The best-fit spectral parameters for all datasets are summarized in Table~\ref{tab:spec}, and the corresponding spectra with model fits are shown in Fig.~\ref{fig:allspec}.

\begin{table*}
	\centering
	\caption{Best-fit spectral parameters of \src\ with partially absorbed \texttt{APEC} model (\texttt{tbabs*tbpcf*APEC}) from the \chandra, 2002 \xmm, and 2017 \xmm\ observations. The errors and limits are quoted at the 90\% significance level.}
	\label{tab:spec}
 	\resizebox{0.8\linewidth}{!}{
	\begin{tabular}{lcccr} 
		\hline
Model	&	Parameters	&	\chandra\	&	2002 \xmm &	2017 \xmm	\\
\hline											
\texttt{Tbabs}	&	$N_H^{\rm ISM}$ ($10^{22}$ cm$^{-2}$)	&	$0.4\pm0.3$	&	$0.7_{-0.4}^{+1.0}$	&	$0.11 \pm 0.08$	\\

\texttt{Tbpcf}	&	$N_H^{\rm Local}$ ($10^{22}$ cm$^{-2}$)	&	$17_{-7}^{+11}$	&	$41_{-15}^{+81}$	&	$12.4_{-2.3}^{+2.8}$	\\
	&	Cov. frac.	&	$0.70_{-0.13}^{+0.07}$	&	$0.78_{-0.23}^{+0.18}$	&	$0.66_{-0.05}^{+0.03}$	\\
    
\texttt{APEC} &	$kT$ (keV)	&	$>11.4$ &	$>1.6$	&	$11.7_{-1.6}^{+7.5}$	\\
	&	Abund	&	\multicolumn{3}{c}{$1^{\rm fixed}$}	\\
	&	Norm ($10^{-3}$)	& $2.1 \pm 0.4$	&	$1.8_{-0.7}^{+15.8}$	&	$1.56 \pm 0.11$	\\										
\texttt{Cons} &	$C_{\rm MOS1}$	&	&	$1.4 \pm 0.2$	&	$1.01 \pm 0.05$	\\
	&	$C_{\rm MOS2}$	&	&	$1.3 \pm 0.2$	&	$0.95 \pm 0.05$	\\
Flux$^a$	&	$F_{0.9-10~\rm keV}$	&	$2.23 \times 10^{-12}$	&	$1.14 \times 10^{-12}$	&	$1.8 \times 10^{-12}$	\\
\hline											
	&	$\chi^2$/dof	&	31.4/33 &	12.8/23 &	223.8/219	\\
		\hline
    \multicolumn{5}{l}{$^a$Unabsorbed flux (in units of \erg) corrected for the Galatic ISM absorption.}\\
	\end{tabular}}
\end{table*}

\subsubsection{Phase-resolved spectroscopy}

We also conducted a phase-resolved spectral analysis of \src. Since the count statistics of these spectra were low, we could not divide the phases better than two. Therefore, with the aim to analyse the pulse phases corresponding to off-peak or on-peak, we extracted spectrum from two phase ranges, 0.0--0.5 and 0.5--1.0, respectively, defined in Fig. \ref{fig:pulseprofile} from each EPIC camera.  

We simultaneously fit the spectra from two phases with the same emission models used for the phase-averaged spectrum. All models were able to describe the phase-resolved spectra well. The phase-resolved spectra of \src\ also showed emission line features, both during the on-peak as well as off-peak of the pulse phase, indicating the presence of ionized material throughout the pulse phase.

\begin{figure}
	\includegraphics[width=\columnwidth]{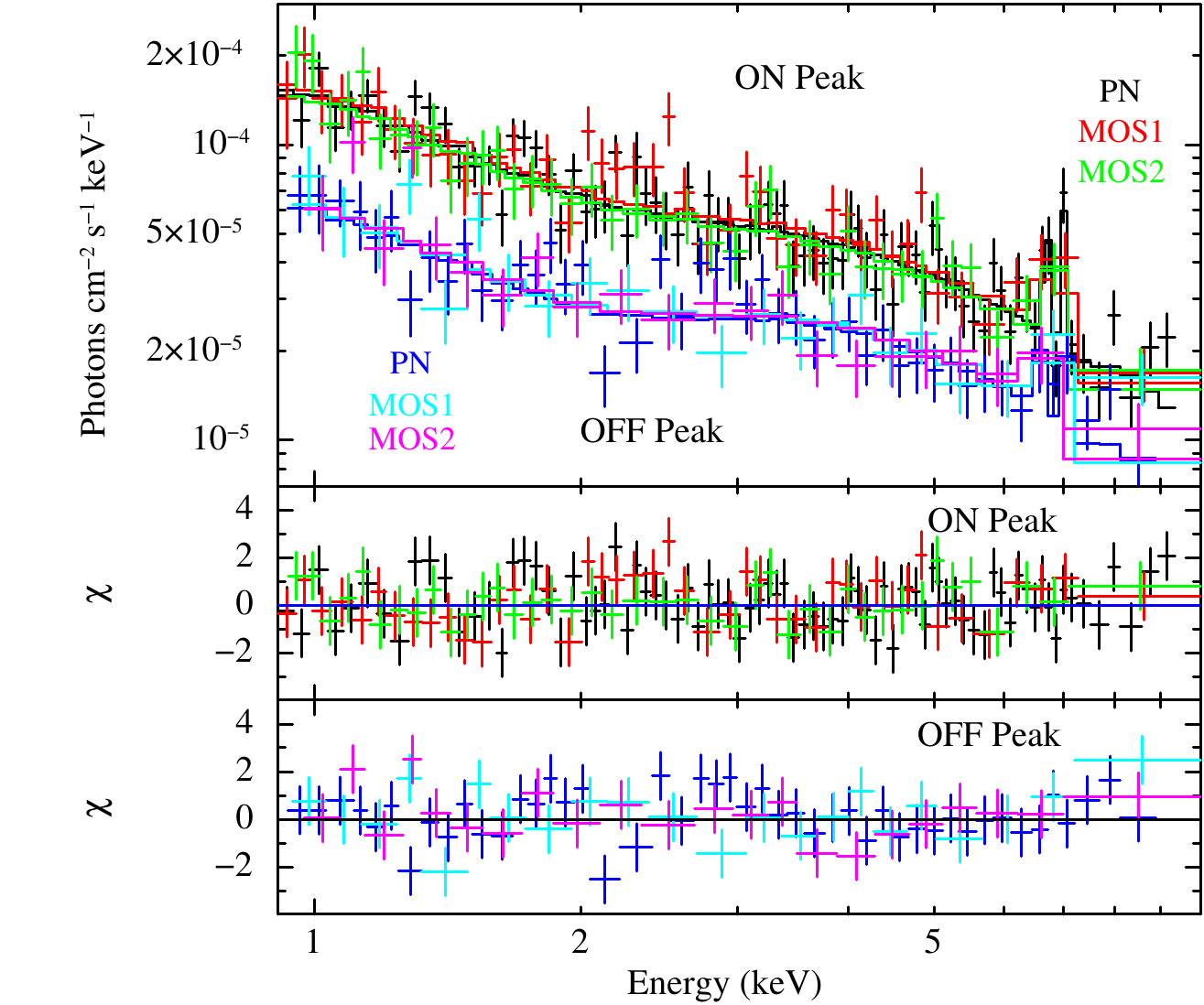}
    \caption{Phase-resolved spectra of \src\ during the on- and off-peak of pulse with the best-fit model (\texttt{tbabs*tbpcf*APEC}). The middle and bottom panels represent the residuals from the best-fit spectral model for the on-peak and off-peak spectra, respectively.}
    \label{fig:spec-phr}
\end{figure}

Here, we focus on the results from the partially absorbed \texttt{APEC} model, which yielded a best-fit $\chi^2$/dof of 251/246. Fig. ~\ref{fig:spec-phr} shows the phase-resolved spectra from on- and off-peak fitted with this model. The results do not provide clear evidence of a spectral variation between the two-phase ranges. The best-fit parameters suggest a consistent plasma temperature of $\sim$12 keV across both phase intervals. Considering the estimated uncertainties, the best-fit values of most parameters are consistent between the two spectra and with the results obtained for the phase-averaged spectrum. However, the partial absorber shows a higher covering fraction for the off-peak spectra ($0.73^{+0.04}_{-0.08}$) compared to on-peak ($0.064^{+0.04}_{-0.10}$). Furthermore, the source flux decreases to $\sim$50\% from $2.31 \times 10^{-12}$ \erg\ during the on-peak to $1.18 \times 10^{-12}$ \erg\ during the off-peak phase, suggesting modulation either due to geometric effects or variable absorption within the accretion structure.

\subsection{Search for Optical and IR Counterparts}

\begin{figure}
\centering
	\includegraphics[width=\columnwidth]{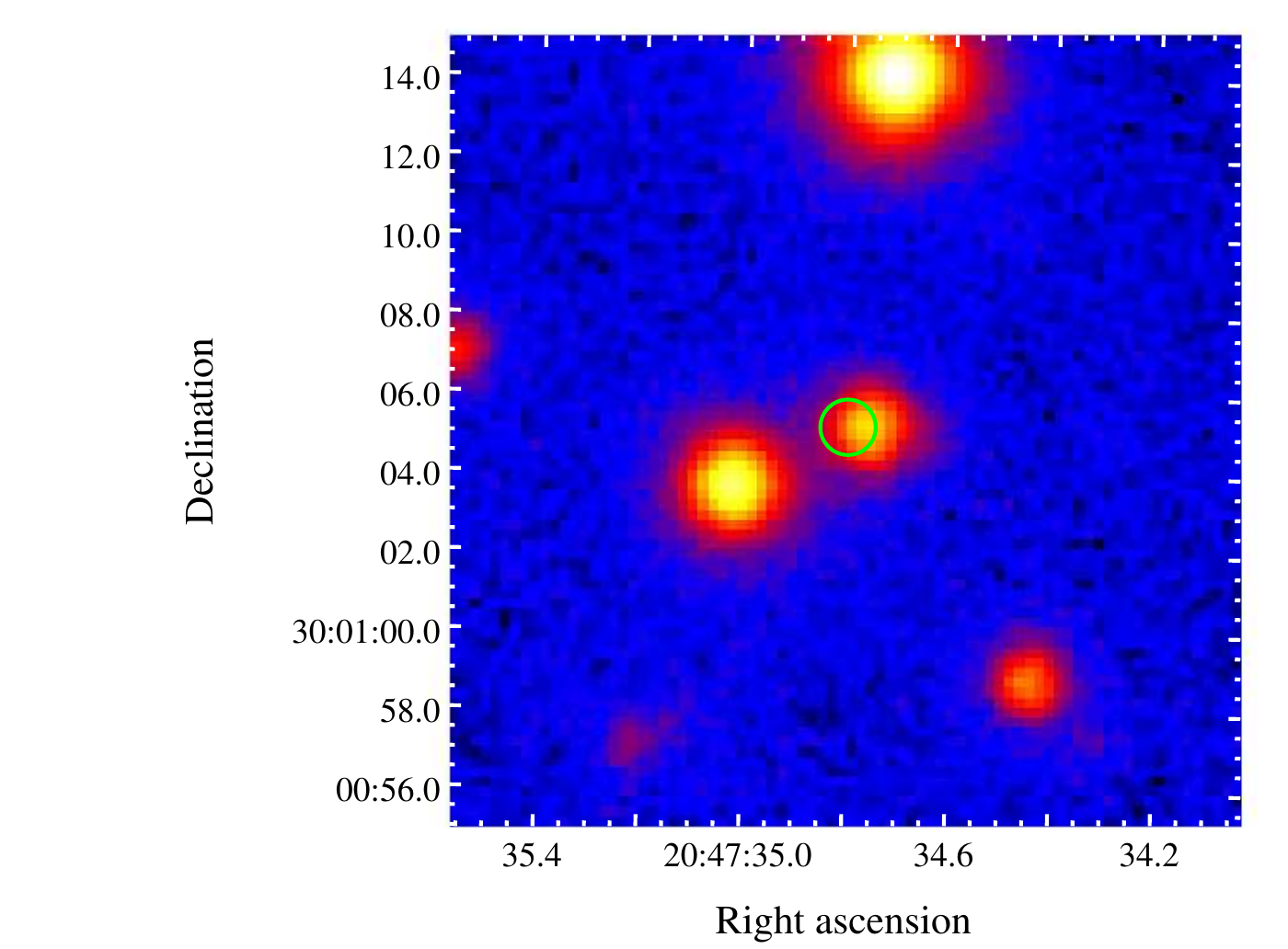}
    \caption{I-band image obtained from the \emph{Pan-STARRS} Archive. The green circle indicates the \chandra\ location of \src\ with a radius of 0.7\arcsec.}
    \label{fig:panstarrs}
\end{figure}

We searched in the archival data and catalogues for the optical and near-IR counterparts of \src\ using the VizieR catalogue access tool \citep{Ochsenbein2000}. In the Two Micron All Sky Survey \citep[\emph{2MASS};][]{Cutri2003, Skrutskie2006}, we identified a source (2MASS 20473475+3001050) within 0.6$''$ of \src\ with $J$, $H$, and $K$ magnitudes of 16.133 (79), 15.558 (108), and 15.00 (12), respectively. The source also matches an entry in the Panoramic Survey Telescope and Rapid Response System (\emph{Pan-STARRS}) survey \citep{Chambers2016} within the 0.5\arcsec. It was detected in the five Pan-STARRS filters $g$, $r$, $i$, $z$ and $y$ with a mean PSF magnitude of 18.4985 (115), 18.2823 (590), 17.8053 (261),  17.7147 (34), and 17.4665 (251), respectively. Fig. \ref{fig:panstarrs} shows the $g$-band image of the optical source marked with a green circle at \chandra\ position with its $1\sigma$ positional uncertainty of 0.7$''$ \citep{Wang2016, Evans2019}. 

Additionally, we identified this source in \emph{Gaia} DR3 \citep{Gaia2023} within 0.5\arcsec\ of \src\ with a $G$-band magnitude of 18.252 (8). Several properties of the companion star were reported based on 1D astrophysical parameters produced by the astrophysical parameter inference system (\textit{Apsis}) processing chain developed in \emph{Gaia} DPAC CU8, assuming it is a star \citep{Gaia2023, Andrae2023}. Parameters reported from General Stellar Parametriser for Photometry (GSP-Phot) using BP/RP spectra include an effective temperature $T_{\rm eff} \sim 6105$ K, surface gravity of $\sim 5\times10^{4}$ cm s$^{-2}$, reddening $E(G_{\rm BP}-G_{\rm RP}) \sim 0.4764$, radius of the star $R\sim0.6447 R_\odot$ and distance to the source of $\sim$2.37 kpc. However, \citet{Gaia-distance2021} reported geometric and photogeometric distances of $\sim$3.96 kpc and $\sim$4.96 kpc, respectively, using parallaxes, colour and apparent magnitude measurements of the star.

\begin{figure*}
\centering
 \includegraphics[width=0.9\linewidth]{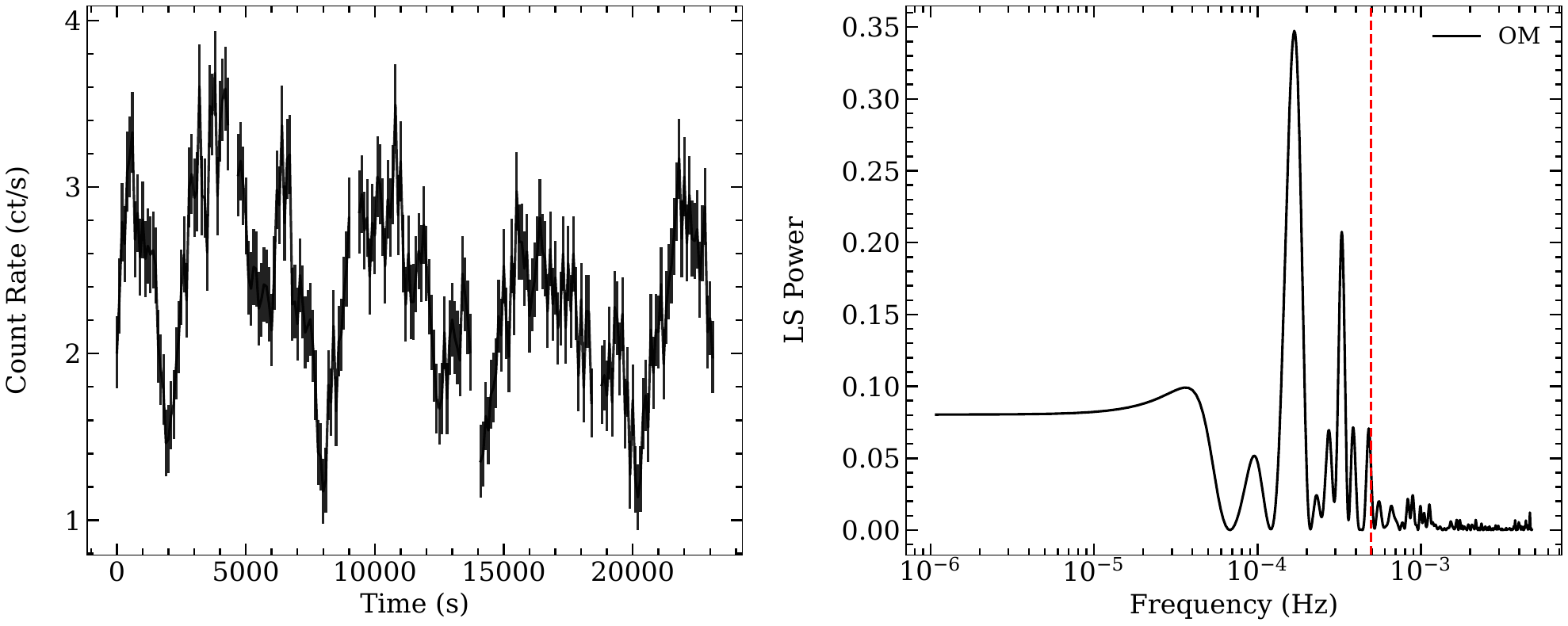} 
    \caption{\textit{Left panel}: The OM light curve of \src\ from 2017 \xmm\ observations binned at 100 s in the U filter. \textit{Right panel}: The LS periodogram of \src\ generated from OM light curve. A significant peak is observed in the periodogram at $1.7\times 10^{-4}$ Hz corresponding to a period of $\sim$5900 s. The vertical dashed line marks the periodicity at 2030 s.}
    \label{fig:omlc}
\end{figure*}

\begin{figure*}
\centering
 \includegraphics[width=0.9\linewidth]{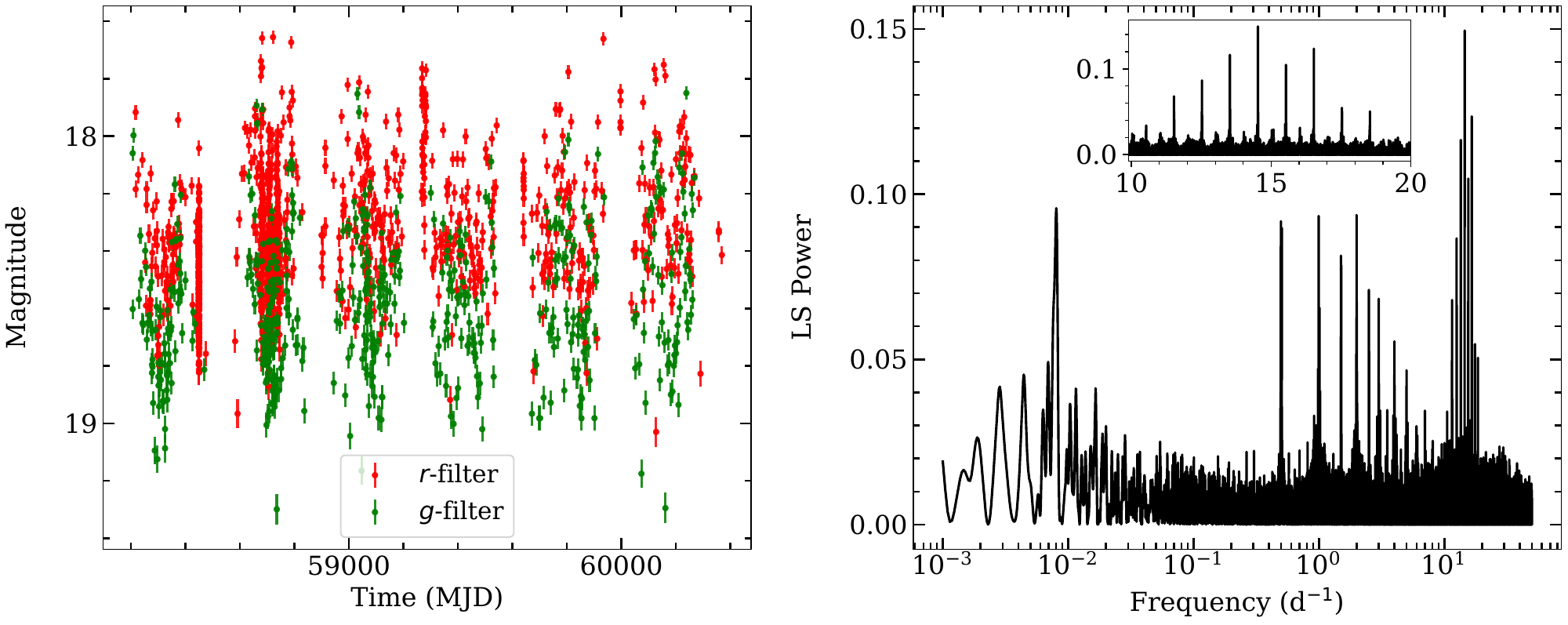} 
    \caption{\textit{Left panel}: The long-term \emph{ZTF} light curve of \src\ in the $g$ and $r$ bands. \textit{Right panel}: The LS periodogram generated from combined data of $g$ and $r$ bands. The inset image shows the rescaled periodogram between frequency 10 and 20 d$^{-1}$, showing the main peak at 14.52 d$^{-1}$. Other peaks are due to the daily aliasing of frequencies in the power spectra.}
    \label{fig:ztf}
\end{figure*}

\begin{figure}
\centering
 \includegraphics[width=0.9\linewidth]{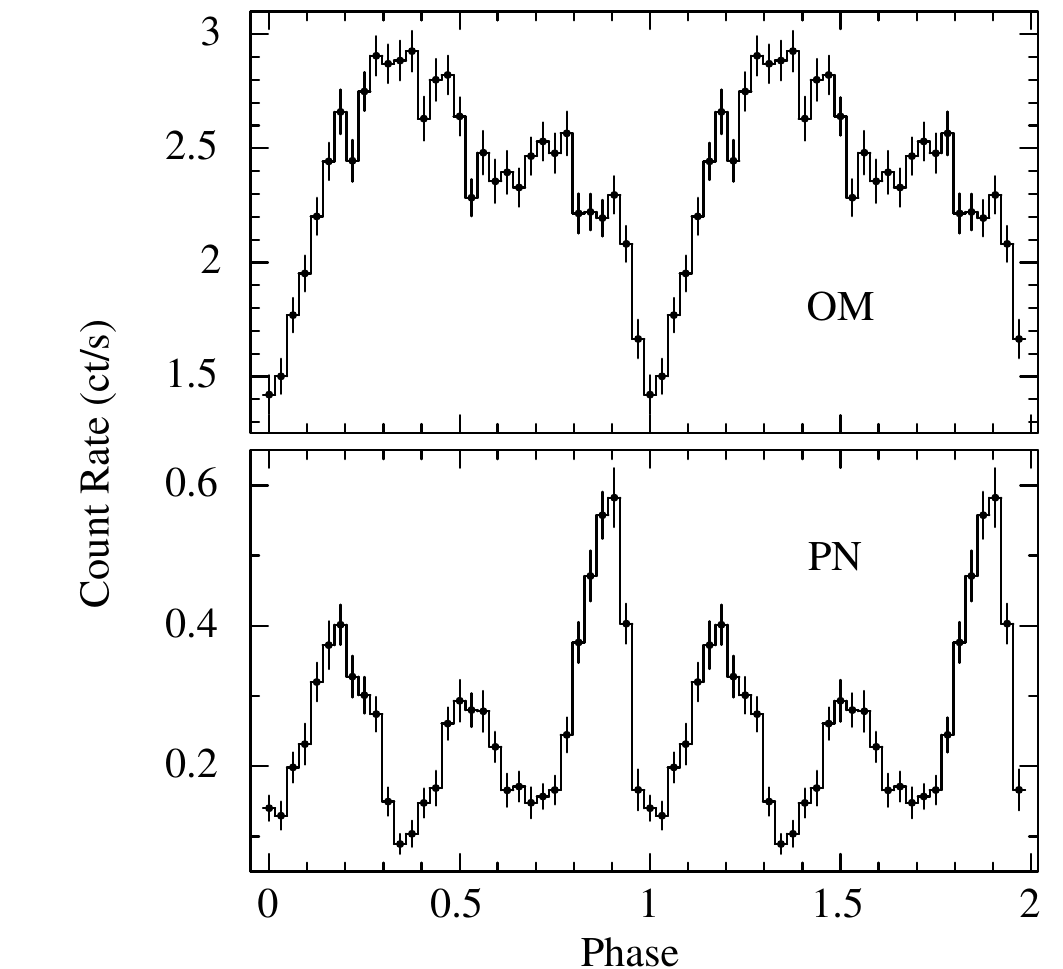} 
    \caption{Folded light curves from the OM (top) and EPIC-PN (bottom) from the 2017 \xmm\ observation, folded at the 5948 s period derived from \emph{ZTF} data.}
    \label{fig:pp-pn-om}
\end{figure}

\subsubsection{XMM-Newton Optical Monitor}

The optical counterpart was also identified in OM. The OM light curve was obtained over a total duration of $\sim$23 ks. Fig. \ref{fig:omlc} illustrates the analysis of the optical data. The left panel shows the light curve, highlighting the temporal variability in the source's emission with an average count rate of $\sim$2.4 counts \psec\ and a standard deviation of about $\sim$0.5 counts. The right panel presents the LS periodogram, which reveals significant periodicity with a primary peak at $1.7 \times 10^{-4}$ Hz, corresponding to a period of $5944 \pm 30$ s with a false alarm probability of $6\times10^{-4}$. Additionally, the second harmonic, at twice the primary frequency, is clearly visible. The periodic signal at 2030 s, previously detected in the EPIC-PN data, is marked with a dashed vertical line and corresponds to the peak at the third harmonic in the OM periodogram. These features in the periodogram represent the dominant periodic signals in the light curve, indicating a characteristic timescale associated with the system's variability in the optical band. 

\subsubsection{\emph{ZTF}}

\citet{Chen2020} reported a source (ZTF J204734.75+ 300104.9) within 0.62$''$ of \src, detected with \emph{ZTF}. This source was reported with mean Palomar/\emph{ZTF} $g$-band and $r$-band magnitudes of 18.544 and 18.334, respectively. It exhibits a suspected periodicity at 1.0041 d in the $g$-band with a false alarm probability of $\sim10^{-3}$. 

To investigate the long-term variability of the source, we utilized \emph{ZTF} observations in the $g$ and $r$ bands. Fig. \ref{fig:ztf} (left) shows the light curves of \src\ in the $g$ and $r$ bands from the \emph{ZTF} DR21 catalogue, spanning $\sim$5.5 years. The source has an average magnitude of $g \approx 18.58$ and $r \approx 18.35$ mag, with weak variability observed at an amplitude of $\sim$0.25 mag. 

To search for periodicity, the LS periodogram was constructed using the combined $g$ and $r$-band light curves. The resulting periodogram, shown in Fig. \ref{fig:ztf} (right), reveals multiple significant peaks, including a prominent power peak at a frequency of $f = 14.52$ d$^{-1}$ (corresponding to a period of $P = 5948$ seconds; see inset figure) with a false alarm probability of $<$$10^{-10}$ and weaker peaks at $n$ and $f \pm n$ (where $n = 1, 2, \dots$ d$^{-1}$), likely arising from modulation by Earth's diurnal rotation. Additionally, a significant low-frequency peak at $8 \times 10^{-3}$ d$^{-1}$ (corresponding to a period of $\sim$124 days) was identified. This long-term variability is likely an artifact of the observation campaign, possibly caused by seasonal data gaps. In CVs, precession timescales are typically much shorter, on the order of only a few days \citep{Montgomery09, Armstrong13}. The periodograms of the individual $g$- and $r$-band light curves similarly display peaks at $8 \times 10^{-3}$ d$^{-1}$, 14.52 d$^{-1}$, and harmonics influenced by Earth's diurnal rotation. However, the significance of the $g$-band peaks is lower, likely due to fewer measurements compared to the $r$-band. Notably, the periodicity of $\sim$1 day reported by \citet{Chen2020} may also be attributed to modulation effects from Earth's diurnal rotation.



\section{Discussion and Conclusions}

In this work, we present a comprehensive spectral and timing analysis of \src, a candidate AM Her-type CV, using data from \chandra\ and \xmm\ observations and optical light curves from OM and \emph{ZTF}. Our analysis reveals several intriguing aspects about its nature and X-ray emission mechanisms. 

The \chandra\ and 2002 \xmm\ spectra showed a relatively hard power-law spectrum compared to the 2017 \xmm\ spectra, which also required additional components (a soft thermal blackbody or partial absorber) to fit the spectra satisfactorily. The 2017 \xmm\ observation also revealed a broad iron emission line, likely due to contributions from both neutral and highly ionized iron, suggesting the presence of ionized material, possibly in the form of an accretion column or disk. All three spectra were satisfactorily modelled using the physically motivated partially absorbed \texttt{APEC} model. The plasma temperature could only be constrained for the 2017 observation to be $\sim$12 keV. The 2002 \xmm\ observation showed a flux approximately 50\% lower than that of the \chandra\ observation, while the 2017 \xmm\ flux was found to be intermediate between the two.

Phase-resolved spectroscopy indicates that the spectral properties remain consistent between the on-peak and off-peak phases, except for variations in the observed flux, which could be due to geometric or intrinsic changes in the emission regions. Iron emission lines were detected in both phases.

Assuming a distance of 2.37 kpc \citep{Gaia2023}, we estimate the X-ray luminosity of the source to be $\sim1.2 \times 10^{33}$ \lum\ in the 0.9--10 keV energy range for the 2017 observation, consistent with typical luminosity $L_X \lesssim 10^{34}$ \lum\ of CVs. 
For the \chandra\ and 2002 \xmm\ observations, the estimated X-ray luminosities fall in the range of $\sim(0.8-1.5) \times 10^{33}$ erg s$^{-1}$.

The X-ray timing analysis of the \chandra\ and \xmm\ data shows significant differences in the observed periodicities and eclipse-like behaviour, challenging the initial classification of \src\ \citep{Israel16}. The \chandra\ observation characterized by a complete 2000-s eclipse-like feature separated by 6000 s, aligns with the expected characteristics of an eclipsing polar-type CV. However, the absence of such apparent eclipses and the detection of a shorter periodicity ($\sim$2000 seconds) in the 2017 \xmm\ observations raise questions about the variability in the system's behaviour. Conversely, optical data from OM (simultaneous with the X-ray) and \emph{ZTF} show a periodicity of around 6000 s. 

The discrepancies observed between X-ray observations taken at different times and accompanying optical observations suggest a dynamic accretion environment in \src, where changes in the accretion rate or geometry over time might lead to varying observational signatures. The $\sim$6000 s periodicity detected in the \chandra\ data, and independently confirmed in the optical band by both the OM and the long-term \emph{ZTF} monitoring, likely represents the true fundamental period of the system. The $\sim$2000 s signal observed in the EPIC X-ray data is therefore plausibly interpreted as the third harmonic of this primary modulation.

Polars are often observed in a low state \citep{Ramsay04}, and their X-ray light curves often consist of alternating bright and faint phases. For example, CW 1103+254 and VV Pup display a faint phase during which X-ray emission is nearly undetectable, attributed to self-occultation of the accretion region (predominantly onto one pole) by the solid body of the white dwarf \citep{Mason85}. In AM Her, a similar faint phase lasting $\sim$0.2 of the orbital cycle is observed, consistent with single-pole accretion interrupted by brief self-occultation \citep{Matt00}. However, high-state observations show complex variation with the disappearance of the faint phase, suggesting a drastic change in the accretion geometry of the source, potentially involving accretion onto a different magnetic pole or multiple poles \citep{Matt00}.

Along these lines, the \chandra\ light curve of \src\ could similarly consist of bright and faint phases, where emission is dominated by one pole with self-occultation by white dwarfs during the faint phase, mimicking an eclipse. During later \xmm\ observations, source emission geometry shifted with a dramatic change in the accretion configuration.  Fig. \ref{fig:pp-pn-om} shows the folded light curves of the OM and EPIC-PN data from the 2017 \xmm\ observation, folded at the 5948 s period identified from the \emph{ZTF} dataset. The X-ray light curve reveals a distinct three-peak structure. Given that AM Her systems often show complex light curves with multiple peaks per cycle, the classification of \src\ as a Polar or AM Her-type system is physically plausible and consistent with structured accretion onto magnetic poles.

However, other possible scenarios may also account for the observed eclipse-like on/off behavior and the discrepancies in periodicity across different observations. These include:

\begin{enumerate}
    \item 
    The possibility of two unresolved transient X-ray sources close to \src. During pointed observations, only one of these sources may be detected, depending on its activity state and brightness at the time \citep[e.g.,][]{Saji16}. This scenario is similar to that observed in the neutron star low mass X-ray binary, GRS 1747--312 \citep{Panurach21, Painter24}. However, there is no clear evidence of contamination.

    \item Two magnetic white dwarfs, FS Aur and HS 2331+3905, exhibit periods that are much longer than their orbital periods, which bear no relation to it, either in light curves or in radial velocity variations measured from spectroscopy \citep{Tovmassian03, Araujo-Betancor05}. These long photometric periods have been associated with the internal regions of the accretion disk \citep{Tovmassian07}. The precession of the rapidly rotating, magnetically accreting white dwarf, consistent with the intermediate polar model, has been proposed as a plausible explanation for such phenomena \citep{Tovmassian07}. Similarly, \src\ may host a moderately magnetized white dwarf accompanied by a geometrically thin accretion disk with two distinct bright spots. The first spot, located on the outer rim of the disk, maintains a fixed position relative to the binary components. The second spot, positioned at the inner edge of the disk, moves in response to the precession of the white dwarf, potentially driving the observed X-ray and optical variability in \src.  
\end{enumerate}


In conclusion, \src\ exhibits complex and variable behaviour that challenges its initial classification as an AM Her-type CV. The differences in periodicity and eclipse-like behaviour between the \chandra\ and \xmm\ observations suggest a dynamic and evolving accretion environment. Nonetheless, the classification of \src\ as an AM Her-type system remains a viable interpretation, considering that these systems commonly produce complex light curves with multiple peaks per cycle due to structured accretion. 
Spectral analysis reveals multiple emission components, including significant iron line emission, indicative of a structured accretion region. The counterparts from \emph{2MASS, Pan-STARRS, Gaia, ZTF}, and OM provide valuable multi-wavelength data that can help in understanding the overall emission mechanisms and the nature of the companion star. Further observations are needed to fully understand the nature and mechanisms driving the observed variability. Continued monitoring and detailed analysis across different wavelengths will be essential in unravelling the mysteries of this intriguing source.


\begin{acknowledgments}
This research has made use of data obtained from the \textit{Chandra Data Archive} and \xmm\ archive. This paper employs a list of Chandra datasets obtained by the Chandra X-ray Observatory contained in ~\dataset[DOI: 10.25574/00740]{https://doi.org/10.25574/00740}. \xmm\ is an ESA science mission with instruments and contributions directly funded by ESA Member States and NASA. This research has made use of the software provided by the \textit{Chandra X-ray Center} (\textit{CXC}) in the application package \textit{CIAO} and of software and tools provided by the High Energy Astrophysics Science Archive Research Center (HEASARC).  The research has also made use of \emph{ZTF} Survey, which is supported by the National Science Foundation under Grants No. AST-1440341 and AST-2034437 and a collaboration including current partners Caltech, IPAC, the Oskar Klein Center at Stockholm University, the University of Maryland, University of California, Berkeley, the University of Wisconsin at Milwaukee, University of Warwick, Ruhr University, Cornell University, Northwestern University and Drexel University. Operations are conducted by COO, IPAC, and UW.
CJ acknowledges the financial assistance received from the ANRF (erstwhile SERB)–DST grant (CRG/2023/000043). We also thank the anonymous referee for insightful comments and suggestions.
\end{acknowledgments}

%

\vspace{5mm}
\facilities{Chandra, XMM-Newton, ZTF}


\software{\textsc{astropy} \citep{astropy}, \textsc{matplotlib} \citep{matplotlib}, the VizieR catalogue access tool, CDS, Strasbourg, France (DOI: 10.26093/cds/vizier), CIAO \citep{Fruscione2006}, XMM-SAS and HEASOFT.}

\bibliography{ref}{}
\bibliographystyle{aasjournal}



\end{document}